\documentclass[prl,aps, showpacs]{revtex4}
\usepackage{dcolumn}
\usepackage{bm}
\begin{document}

\title{Independence on the Abelian Projection of Monopole Condensation in QCD}

\author{A. Di Giacomo}
\affiliation{%
Dipartimento di Fisica dell'Universita', \\
Via Buonarroti 2 56127 Pisa, Italy , \\
  and  INFN Sezione di Pisa}
\email{adriano.digiacomo@df.unipi.it}

\begin{abstract}
   It is proven that dual superconductivity of QCD vacuum in the confining
phase is an intrinsic property, independent on the choice of the abelian
projection used to define the monopoles.
\end{abstract}
\maketitle

Dual superconductivity of the vacuum was proposed in the early times
of QCD as a mechanism for color confinement\cite{1,2}. Here dual means
interchange of electric and magnetic with respect to ordinary
superconductivity. The basic idea is that, in the confining phase,
magnetic charges condense in the vacuum, and the chromoelectric field
acting between two colored particles is channeled into Abrikosov flux
tubes by dual Meissner effect, giving an energy proportional to the
distance, $ V(r) = \sigma  r$, or confinement. Above the deconfining transition
this phenomenon should disappear.

   When trying to make this idea more precise, however, it looks less
unique than expected. A procedure known as Abelian Projection\cite{3}  is
needed to define monopoles, which implies the choice of an operator
  $\Phi(x) = T^a \Phi^a (x)$ , transforming in the adjoint 
representation of the gauge
group, say SU(N) . After diagonalization of  $\Phi(x)$    in color space by a
gauge transformation (Abelian Projection), $N-1$ $U(1)$ residual gauge
symmetries exist, whose magnetic charges are conserved. Dual
superconductivity means Higgs breaking of these magnetic $U(1)$ symmetries.
A priori monopoles defined  by different abelian projections, i.e. by
different choices of the operator $\Phi$, are not related to each other. Indeed
their number and locations are different when observed in any given field
configuration. The question is then what monopoles do condense in the
vacuum to produce dual superconductivity. One possibility is that some
abelian projection is privileged (see ref\cite{4} for a review). Another
extreme possibility is that all abelian projections are physically
equivalent, as guessed in ref\cite{3}.

    The mechanism of confinement by dual superconductivity has been
investigated on the lattice\cite{5,6,7}, by studying the expectation
value of an operator  $\mu (x)$ , which creates a monopole in some abelian
projection. The expectation is that in the confined phase ( Higgs phase)
its vev  $\langle\mu\rangle$ is non-zero signalling monopole 
condensation; in the
deconfining phase   $\langle\mu\rangle   = 0$,the Hilbert space being 
superselected with
respect to magnetic charge. Around $T_c$,  $\langle\mu\rangle_{T\to 
T_c}=(1 - T/T_c)^\delta$ .
$\langle\mu\rangle$ is called a
disorder parameter in the language of statistical mechanics, being the
order parameter of the strong coupling (disordered) confining phase.
   What is found by numerical simulations is that indeed 
$\langle\mu\rangle\neq0$        at
$T < T_c$ ,  $\langle\mu\rangle   =0$ at  $T > T_c$,  $\delta = 
.20(3)$               for SU(2),
$\delta = .50(3)$ for SU(3), and that $\langle\mu\rangle$  is 
independent of the choice of
the abelian projection\cite{5,6,7} within errors.

In this letter we show that the independence of  $\langle\mu\rangle$ 
on the abelian
projection follows from gauge invariance, and that it also
holds for any correlator $\langle T [\mu(x_1)\ldots\mu(x_n)] \rangle 
$   of creation operators
of monopoles. It follows that dual superconductivity of the confining
phase is an intrinsic property, independent of the choice of the abelian
projection, and so is the coulomb nature of the deconfined phase.

The key quantity to define the abelian projection corresponding to an
operator    $\Phi(x)$ in the adjoint representation is 'tHooft 's 
field strength tensor\cite{7h}
\begin{equation}
F_{\mu\nu} = Tr\left\{\Phi G_{\mu\nu}\right\}- \frac{i}{g}
Tr\left\{ \Phi\left[D_\mu\Phi,D_\nu\Phi\right]\right\}\label{eq1}
\end{equation}
  $F_{\mu\nu}$is gauge invariant and color singlet. The notation is standard
\[ \Phi = \Phi^a T^a,\quad A_\mu = a_\mu^a T^a,\quad G_{\mu\nu} =
\partial_\mu A_\nu - \partial_\nu A_\mu + i g \left[A_\mu,A_\nu\right],
\quad D_\mu\Phi = \partial_\mu\Phi - i g \left[A_\mu,\Phi\right]\]
$T^a$ are the group generators, with $Tr\{T^a T^b\} = \delta^{ab} $. 
For particular choices of
  $\Phi$   the tensor in eq(1) becomes an abelian field strength in the gauge in
which  $\Phi$   is diagonal.

For $SU(2)$ gauge group this happens for any operator $\hat\Phi = 
\vec\varphi\cdot\vec T$ with
$\vec\varphi^2    = 1$. In that case indeed\cite{7g} bilinear terms 
in $ A_\mu  A_\nu$  cancel
between the two terms of eq(1), and
\begin{equation}
F_{\mu\nu} = Tr\left\{\partial_\mu(\hat\Phi a_\nu) - 
\partial_\nu(\hat\Phi A_\mu)\right\}
-\frac{i}{g}Tr\left\{ 
\hat\Phi\left[\partial_\mu\hat\Phi,\partial_\nu\hat\Phi\right]\right\}
\label{eq2}\end{equation}
In the gauge in which  $\hat\Phi = \hat\Phi_{diag}$, or $\vec\varphi 
= (0,0,1)$, $\partial_\mu\varphi = 0$                       and
\begin{equation}
F_{\mu\nu} =\partial_\mu A^3_\nu - \partial_\nu A^3_\mu
\label{eq3}\end{equation}
The generic operator $\Phi$ can be written $\Phi = c(x) \hat\Phi(x)$  with
\begin{equation}
\hat\Phi(x) = U^\dagger(x)\Phi_{diag} U(x)
\label{eq4}\end{equation}
$U(x)$ is the abelian projection, and the residual $U(1)$ has $T^3$ as a
generator in the abelian projected gauge.

   The operator which creates a monopole at  $(\vec x ,x^0 )$ is\cite{5,7a}
\begin{equation}
\mu(\vec x ,x^0 ) = \exp\left(
\int d^3y \vec b_\perp(\vec x - \vec y)\cdot Tr\left\{\vec E(\vec 
y,x^0) \hat\Phi(x)\right\}\right)
\label{eq5}\end{equation}
where  $\vec E$  is the electric field operator,and $\vec 
b_\perp(\vec x-\vec y)$ is the classical
vector potential produced at  $\vec y$  by a monopole sitting at 
$\vec x$, in the
transverse gauge $\vec\nabla b_\perp(\vec x-\vec y) = 0$. The 
operator defined by eq.(5) is
nothing but the translation of the field  $\vec A^3_\perp(\vec y, 
x^0)$ by  $\vec b_\perp(\vec x-\vec y)$
in the
abelian projected gauge. In the Schroedinger picture
\begin{equation}
\mu | \vec A^3_\perp(\vec y, x^0)\rangle =
| \vec A^3_\perp(\vec y, x^0) + \vec b_\perp(\vec x-\vec y) \rangle
\label{eq6}\end{equation}
since $\vec E^3_\perp$ is the conjugate momentum to $\vec A^3_\perp$. 
More generally
any configuration of monopoles and antimonopoles can be added at $t=x^0$
by an appropriate choice of $\vec b_\perp$. By use of eq.(4) and of 
the cyclicity
of the trace the operator $\mu$  can be written
\begin{equation}
\mu(\vec x ,x^0 ) = \exp\left(
\int d^3y \vec b_\perp(\vec x - \vec y)\cdot Tr\left\{U \vec E(\vec 
y,x^0) U^\dagger\hat\Phi_{diag}\right\}\right)
\label{eq7}\end{equation}
The unitary transformation  $U(x)$ is what distinguishes one abelian
projection from another.

   The correlator of any number of monopole fields is given by
\begin{equation}
\langle T [\mu_1(x_1)\ldots\mu_n(x_n)] \rangle =
\frac{1}{Z} \int dM \mu_1(x_1)\ldots\mu_n(x_n)
\label{eq8}\end{equation}
$Z = \int dM$. A change of variables $A_\mu  \to  U A_\mu U^\dagger+ 
i \partial_\mu U U^\dagger$ leaves the measure $dM$
invariant and replaces $U E U^\dagger$  by  $E$  in the expression 
for $\langle\mu\rangle$ eq(7).
This proves that the correlator eq.(8) is projection independent:
whatever the choice of $U$   all the  $\mu$'s can be replaced by
\begin{equation}
\mu(\vec x ,x^0 ) = \exp\left(
\int d^3y \vec b_\perp(\vec x - \vec y)\cdot Tr\left\{ \vec E(\vec 
y,x^0)\hat\Phi_{diag}\right\}\right)
\label{eq9}\end{equation}
which is projection independent. In particular  $\langle\mu\rangle$ 
is projection
independent, and such is the Higgs nature of the confined phase and the
Coulomb nature of the deconfined one.

   For generic  $SU(N)$  the argument is analogous. The choices for $\Phi$  in
eq.(1), which lead to eq.(2) can be identified\cite{8}:  there are  $N -1$
independent possibilities  $\Phi^a = U^\dagger \Phi^a_{diag} U$  with 
arbitrary $U(x)$
and
\begin{equation}
\Phi^a_{diag} = \left(
\overbrace{\frac{a}{N},\frac{a}{N},\ldots,\frac{a}{N}}^{b},
\overbrace{-\frac{b}{N},-\frac{b}{N},\ldots,-\frac{b}{N}}^{a}\right)\qquad 
b = N-a
\label{eq10}\end{equation}
Any operator  $\Phi$     in the adjoint representation is of the form
  $\Phi = U^\dagger\Phi_{diag} U$  with $ U $ a gauge  transformation. 
Since the $\Phi^a_{diag}$ are a
complete basis for diagonal matrices $\Phi = \sum c_a U^\dagger 
\Phi^a_{diag} U$.
   In the abelian projected gauge the field  $F^a_{\mu\nu}$ 
corresponding to a given
   $\Phi^a$    will be
\begin{equation}
F^a_{\mu\nu} = Tr\left\{\partial_\mu(\Phi^a_{diag} A_\nu) - 
\partial_\nu(\Phi^a_{diag} A_\mu)\right\}
\label{eq11}\end{equation}
    Only the diagonal part of  $A^\mu$  contributes. $A^\mu_{diag}$ 
can be expanded in
terms of a complete set of diagonal matrices $\alpha^a$, which obey the
orthonormality relations  $Tr(\Phi^a \alpha^b) = \delta^{ab}$,
\begin{equation}
\alpha^a_{diag} = (\overbrace{0\ldots 0}^{a-1},1,-1,0,\ldots 0 ) 
\label{eq12}\end{equation}
The $N - 1$ abelian fields produced by the abelian projection correspond to
the generators eq.(12). The operator which creates a monopole of the
a-th $U(1)$ is then
\begin{equation}
\mu^a(\vec x ,x^0 ) = \exp\left(
\int d^3y \vec b_\perp(\vec x - \vec y)\cdot 
Tr\left\{U^\dagger\Phi^a_{diag} U \vec E(\vec y,x^0) \right\}\right)
\label{eq13}\end{equation}
or
\begin{equation}
\langle\mu^a\rangle = \langle\exp\left(
\int d^3y \vec b_\perp(\vec x - \vec y)\cdot Tr\left\{\Phi^a_{diag} U 
\vec E(\vec y,x^0)
U^\dagger
\right\}\right)\rangle
\label{eq14}\end{equation}
   Again in the correlator of any number of $\mu$'s    the unitary 
transformation
$U$  can be reabsorbed  by a gauge transformation which leaves the measure
invariant,proving the independence on $U$, i.e. on the abelian projection.

   All that can be trivially translated into the regularized version of the
theory on the lattice\cite{5,6,7}. In fact in ref.\cite{7} a numerical
comparison was done between  $\langle\mu\rangle$    corresponding to 
a number of abelian
projections, and the $\langle\mu\rangle$   of the form eq.(14), 
considered as an average on
abelian projections, and they were equal within errors: we can now say
that this only means that our numerical procedure was correct.

  In conclusion the correlators of monopole fields are abelian projection
independent, and such should be the field theory reconstructed from them.
In particular dual superconductivity of the vacuum is an
intrinsic,projection independent, property.

\begin{acknowledgments}
   Thanks are due to M.D'Elia. L. Del Debbio and to G.Paffuti for 
helpful discussions.
This work was partially supported by MIUR project Teoria e fenomenologia
delle particelle elementari.
\end{acknowledgments}

\end{document}